# DSSSD based Spectrometer of the Dubna Gas-Filled Recoil Separator

Yury S. Tsyganov

*FLNR, JINR 141980 Dubna, Moscow Region, Russian federation*

*Abstract*–**Recently, in the DGFRS related experiments the discoveries of new superheavy elements Z=114-118 were successfully performed. As concerns to a background radical suppression in those experiments it was namely "active correlations" method which was applied to this goal. Note, that application of DSSSD detector allows one to apply this method more effectively, although with some difficulties due to neighbor strips edge effects. Present version of the DGFRS spectrometer is presented. Shapes of ER's registered energy spectra in the $^{nat}$Yb+$^{48}$Ca reaction are presented. First results of its application in the $^{240}$Pu+$^{48}$Ca→Fl\* complete fusion nuclear reaction are presented too. Conclusion, that edge effects are to a first approximation explained by the role of inter strips area is drawn.**

## I. INTRODUCTION

THE existence of superheavy elements (SHE) was predicted in the late 1960-s as one of the first outcomes of the macroscopic-microscopic theory of atomic nucleus. Modern theoretical approaches confirm this concept. To date, nuclei associated with the "island of stability" can be accessed preferentially in $^{48}$Ca-induced complete fusion nuclear reactions with actinide targets. Successful use of these reactions was pioneered employing the Dubna Gas-Filled Recoil Separator (DGFRS) [1] at the Flerov Laboratory of Nuclear Reactions (FLNR) in Dubna, Russia. In the last two decades intense research in SHE synthesis has taken place and lead to significant progress in methods of detecting rare alpha decays. Method of "active correlations" used to provide a deep suppression of background products is one of them. Significant progress in the detection technique was achieved with application of DSSSD detectors. Note that applying the method of "active correlations" with DSSSD detector is even more effective compared with the case of resistive PIPS detector. On the other hand, some specific effects take place and possible sharing registered signal between two neighbor strips from p-n junction side is one of them.

## II. DETECTION MODULE OF THE DGFRS: PRESENT STATUS

The DGFRS is one of the most effective facilities in use for the synthesis of SHE. Using this facility it has been possible to obtain more than fifty new superheavy nuclides. In long-term experiments aimed to the synthesis of SHE one should take into account that yield of the products under investigation is small enough, usually – one per days – one per month, thus the role of the detection system and focal plane detector is quite significant as well as beam intensity requirements. Since 2015, to increase the position granularity of the detectors, which reduces the probability of observing sequences of random events that could be imitate decay chains of synthesized nuclei, the new focal plane detector has been used. It consists of 120x60 mm$^2$ 48x128 strips Micron Semiconductor **D**ouble **S**ide **S**ilicon **S**trip **D**etector (DSSSD). Design of this detector and CAMAC spectrometer of the DGFRS are shown in the Fig.1a,b.

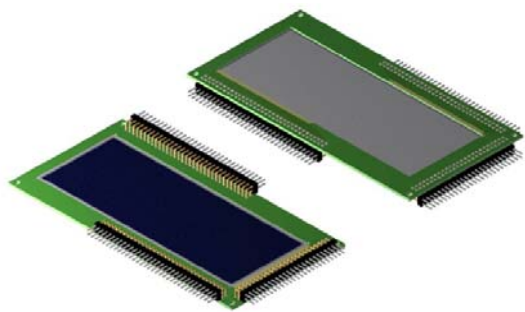

Fig. 1a. DSSSD 48x128 strips focal plane detector of the Dubna Gas-Filled Recoil Separator spectrometer

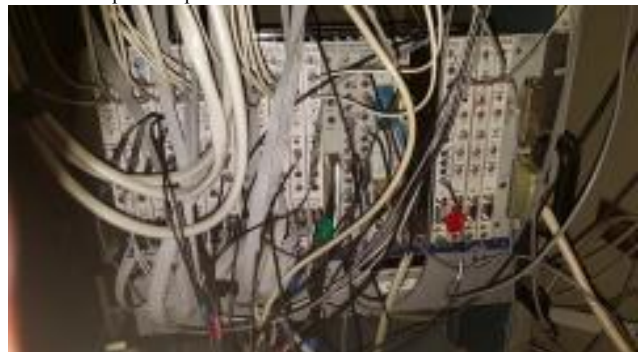

Fig. 1b. View of the DGFRS CAMAC based spectrometer

The detection system of the DGFRS was calibrated by registering the recoil nuclei and decays (α, SF) of known isotopes of No and Th and their descendants produced in the reactions $^{206}$Pb($^{48}$Ca,2n) and $^{nat}$Yb($^{48}$Ca,3-5n), respectively. Before implantation into the focal plane detector, the separated ERs passed through a time-of-flight (TOF) measuring system that consists of two (start and stop) multiwire proportional chambers filled with pentane at ≈1.6 Torr [2]. The TOF system allows to distinguishing recoils coming from the separator and passing through the TOF system from signals, arising from α decay or SF of the implanted nuclei (without TOF or ΔE1 or ΔE2 signals). In order to eliminate the background from the fast light charged particles (protons, α's, etc produced from direct reactions of projectiles with the DFFRS media) with signal amplitudes lower than registration threshold of the TOF detector, a "VETO" silicon detector is placed behind the front detector. From the theoretical calculations and the available experimental data, one can estimate the expected α-particle energies of the ERs and their descendant nuclei that could be produced in a specific heavy-ion induced reaction of synthesis. For α particles emitted by the parent or daughter nuclei, it is possible to chose wide enough energy and time gates $\Delta E_{\alpha 1}$, $\Delta t_{\alpha 1}$, $\Delta E_{\alpha 2}$, $\Delta t_{\alpha 2}$ etc. and to employ a specific low-background detection scheme – method of "active correlations".

### III. METHOD OF "ACTIVE CORRELATIONS"

The simple, but very effective idea of the mentioned method is as following. PC-based Builder C++ program is aimed at searching in real-time mode of time-energy-position recoil-alpha links, using the two matrix representation of the DSSSD detector separately for ER matrix and α-matrix. In each case of "alpha particle" signal detection, a comparison with "recoil (ER)" matrix is made. If the elapsed time difference between "recoil" and "alpha particle" within preset time value, the system turns on the cyclotron beam chopper which deflects the heavy ion beam in the injection line of the U-400 FLNR cyclotron for a definite time interval (usually 0.5-2 min). The next step of the computer code ignores horizontal position (128 strips from p-n junction side) of the forthcoming alpha-particle signal during the beam-off interval. If such decay taken place in the same vertical position strip (48 strips) that generated the pause, the duration of the beam-off interval is prolonged by a factor 5-10. The dead time of the system, associated with interrupting the beam is about 110 μs, including linear growth chopper operation delay (~10 μs) and estimated heavy ion orbit life-time (~60μs). In contrast to former resistive layer PIPS detector application [3,4], using of DSSSD detector one has three main specific features:

1. ER matrix (48x128 elements) de-facto already exists due to discrete composition of the DSSSD detector;
2. On the other hand, edge effects between the neighbor p-n junction side strips should be taken into account (128 strips in our case);
3. From the viewpoint of radiation durability off DSSSD it should be mentioned that detector is operated strongly it total depletion mode.

New version of software, reported below, takes into account points 1 and 2.

#### A. GNS-2016 Builder C++ program package

GNS-2016 Builder C++ program package has been designed to work together with new DSSSD based detection module of the DGFRS and appropriate electronics. It consists of two main parts:
- ERAS-2016.exe – data taker and file writer also used to generate beam stop signal;
- MONITOR-2016.exe – a visualization unit also used for exact tuning of TOF-ΔE low pressure, pentane filled module;
- Some programs used for testing electronics modules are also within this package.

#### B. ERAS – 2016 Builder C++ data taking program

ERAS-2016 C++ program (**ER** –**A**lpha **S**equences) is designed to provide data taking, file writing and to search for ER-α correlated sequences in a real-time mode. The block-diagram of this process and the flow chart of the program are shown in the Fig.2 a,b, respectively. Note, that beam is chopped in the cyclotron injection line, when the value of $^{48}$Ca projectile energy is small enough ($^{48}$Ca beam energy is ~18 kV at the position of the beam chopper). The code brunch, that is responsible for real-time search for ER-α sequence is shown in gray in Fig.2b.

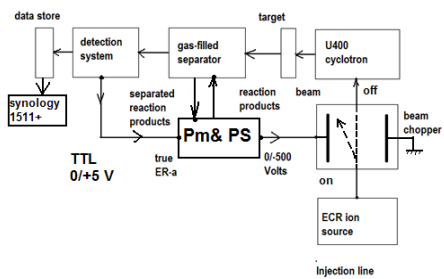

Fig. 2a,. Block-diagram of the process to search for ER-alpha chains and to provide beam stops. (Pm& PS – parameter monitoring and protection system of the DGFRS)

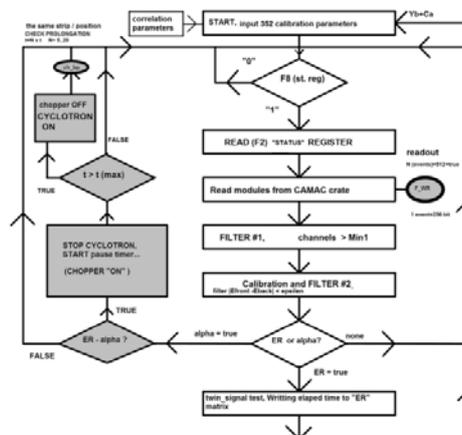

Fig. 2b,. The flow chart of the ERAS C++ program. Brunch for searching for ER-alpha sequences is shown in gray in the left picture side. Calibration parameters are extracted from $^{nat}$Yb+$^{48}$Ca →Th* nuclear reaction (352 constants)

ERAS correlation parameter list is represented below:
- ER-α correlation time to provide a beam stop;
- Integer value (5-20) which denotes that in the case of prolongation a beam-off interval, the pause will be a factor 5 to 20 longer;
- Minimum and maximum values of ER and α-particle signals set to stop the beam and min and max value of alpha particle signal set for prolongation of beam-off interval;
- Minimum and maximum values of TOF and $\Delta E_{1,2}$ signals measured with low pressure gaseous TOF module.

Routines "Filter#1" and "Filter#2" shown in Fig.2b provide filtering of incoming signals according to channel number and energy, respectively. The routine "check prolongation" is active only when the beam chopper is in "switch on" state, otherwise it provides no extra operation. Distinguishing between true/false Boolean "twin signal" variable is performed by reading of the appropriate eight bits of "status" CAMAC 1M register unit. Each bit in "1" state corresponds to operation of 16-input analog-to-digital converters (ADC). Additionally, ERAS program generates text file with parameters of every beam stop. It includes energy signals of recoil and alpha particle from both front and back strip of the DSSSD, elapsed time of the ER signal and time difference between alpha particle signal and ER (recoil) signal, numbers of and one bit marker (0/1) indicating simultaneous operation of two neighboring strips on p-n junction side. In $^{251,249}Cf+^{48}Ca$ reaction experiment at beam intensity ~0.7 pμA, such "double" events from DSSSD back side strips amounted to 15.9% of total number. In this case, the program calculates actual back strip energy in the form:

$E_{back}=a_i N_i + b_i + a_{i+/-1} N_{i+/-1} + b_{i+/-1}$. Here, $(a_i, b_i)$-calibration constants, i=1..128.

### IV. MONITOR-2016 C++ CODE FOR FILE PROCESSING

C++ Builder MONITOR-2016 program is designed for processing of files generated by ERAS program. The program constructs spectra for each front and back strip and for $\Delta E$ and TOF signals (totally, 250 histograms). Except for building histograms, some specific spectra are built by the program. For example, it provides output files constructed as sum alpha spectra meeting a condition:
a) all signals TOF=0 and $\Delta E_{1,2}=0$;
b) the same as a) condition, but additionally, single-bit flight marker is equal to zero.

This flight marker is generated if at least one signal from start or stop gaseous counter exceeds a 40-mV threshold of an one-shot unit; in this case, the latter generates 0/+5 V output TTL signal with duration about 20 μs (preamplifier response to typical ER signal is ~0.5-1 V and about ~ 50 mV for 5.5 MeV alpha particles). Of course, with low-threshold one-shot unit, certain precautions must be made in order to avoid extra suppression of true α-particle signals of implanted nuclei decays.

### V. EXAMPLE OF APPLICATION OF BUILDER C++ ERAS CODE IN THE $^{240}Pu+^{48}Ca \rightarrow Fl^*$ COMPLETE FUSION NUCLEAR REACTION

In the long term $^{240}Pu+^{48}Ca\rightarrow Fl^*$ experiment the beam was interrupted after the detection of recoil signal with the expected implantation energy for Z=114 evaporation residues followed by an α-like signal in the front detector with the energy 9.8 – 11.5 MeV, in the same (or neighbor) DSSSD pixel. The ER energy interval was chosen to be 6 – 16 MeV. The triggering ER-α time interval was set to 1 s. The beam off interval was set 1 min. In this time, if an α-particle with Eα 8.5 to 11.5 MeV was registered in the same front strip as the ER signal, the beam off interval was automatically extended to 5 min. During the experiment, two chains were detected that were attributed to Z=114 nuclei [5]. These are presented in the Fig.3. The registered ER energy amplitudes are shown in the Fig.4 and are in a good agreement with the theoretical calculation [6].

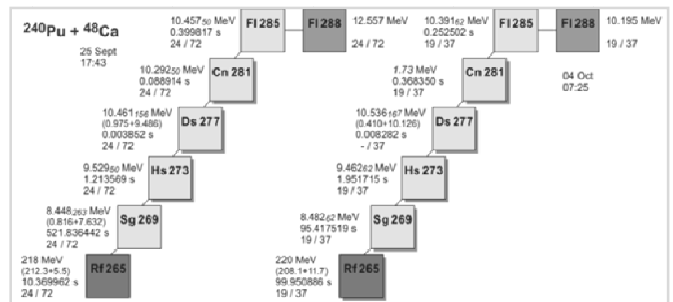

Fig. 3 Two chains of Z=114 nuclei decay detected in the $^{240}Pu+^{48}Ca$ experiment

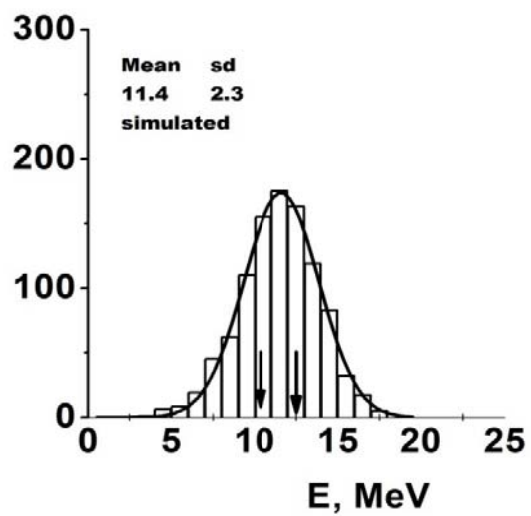

Fig. 4 Two ER registered energy events of Z=114 nuclei detected in the $^{240}Pu+^{48}Ca$ reaction. Left upper corner – results of Gaussian fit of computer simulation

## VI. SUMMARY

- Together with the higher granularity advantage, using of DSSSD detector arises some local problems, the edge effect between neighbor strips on p-n junction side being one of them. With Borland's Builder C++ GNS-2016 program package this problem was solved. The "active correlations" method was successfully applied in the $^{240}$Pu+$^{48}$Ca→Fl$^*$ experiment using DSSSD based spectrometer of the DGFRS. Measured by the DGFRS DSSSD detector, average ER's energy is in a good agreement with the value calculated one.

For our future projects, associated with putting into operation in 2017 of a new DC-280 ultra intense FLNR cyclotron, we plan to develop more sophisticated algorithms for searching for recoil-alpha or even recoil-alpha-alpha sequences in a real-time mode.

APPENDIX: ISOTOPE $^{217}$TH ALPHA-DECAY CASE OF TWO STRIPS OPERATION

Nuclear reaction $^{nat}$Yb+$^{48}$Ca→$^{217}$Th+3n is very useful for calibration procedure due to a relatively short live time of this thorium isotope. Therefore it is easy to extract ER-alpha correlated chains from the whole data flow. Additionally, this test reaction one can use to study upper described edge effect between two neighbor strips. In the Fig.5 two dimensional picture $E_2 = F(E_1)$ is shown. Here $E_{1,2}$ – energies for any first and second strip, respectively. It can be easily seen that the sum of $E_1+E_2$ is close enough to the alpha decay energy of $^{217}$Th isotope. In the Fig.6a the spectrum for one signal (from two) is presented. To a first approximation, small decreasing in the spectrum middle can be interpreted as a ballistic deficit when charge collection process in the inter strip area (100μm) takes place. In the Fig.6b,c $^{217}$Th recoil registered energy spectrum is shown. In the Fig.7 dependence of back strip measured alpha decay energy against the one measured with front strips is shown.

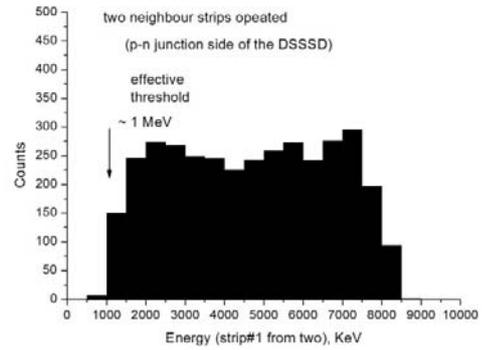

Fig.6a Spectrum of one (from two) $^{217}$Th alpha decay signal

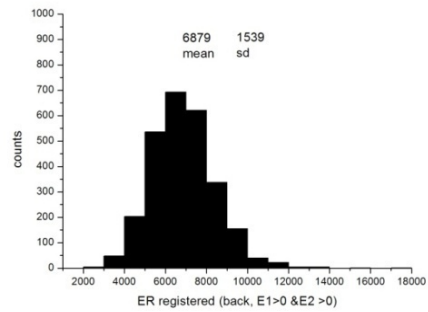

Fig.6b ER's $^{217}$Th registered energy spectrum. Both values $E_1 > 0$ and $E_2 > 0$

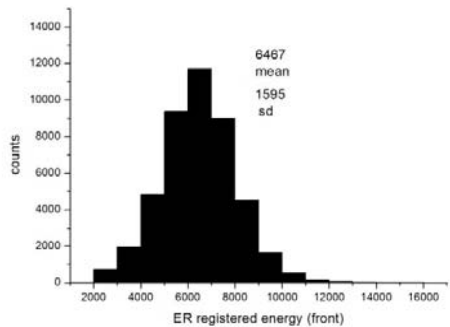

Fig.6c The same as b), but for all recoils $^{217}$Th.

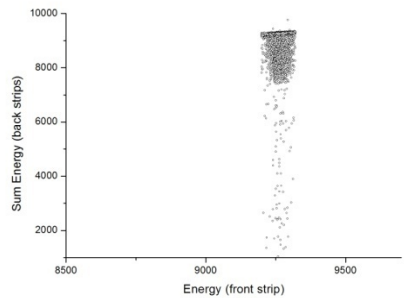

Fig.7 The dependence of sum energy (p-n junction side) against the one measured with front strips

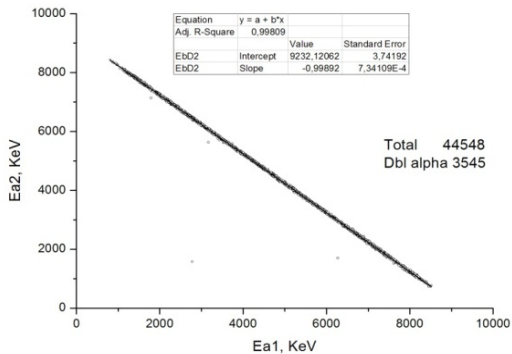

Fig.5 Two dimensional histogram for the case of two signal detection (neighbor strips). Reaction $^{nat}$Yb+$^{48}$Ca→$^{217}$Th+3n

ACKNOWLEDGMENT

This paper is supported in part by the RFBR Grant №16-52-55002/16. The author ise indebted to Dr.s A. Polyakov, M.Shumeiko, V.Zhuchko for their help and fruitful discussions.